\documentclass[twocolumn,amssymb,amsmath,superscriptaddress,aps,pre]{revtex4-1}
\usepackage{epsfig}
\usepackage{times}
\usepackage{bm}

\begin{document}

\title{Activation process in excitable systems with multiple noise sources: Large number of units}

\author{Igor Franovi\'c}
\email{franovic@ipb.ac.rs}
\affiliation{Scientific Computing Laboratory, Institute of Physics, University of Belgrade, P.O. Box 68, 11080 Beograd-Zemun, Serbia}

\author{Matja\v{z} Perc}
\email{matjaz.perc@uni-mb.si}
\affiliation{Faculty of Natural Sciences and Mathematics, University of Maribor,\\
Koro\v{s}ka Cesta 160, SI-2000 Maribor, Slovenia}
\affiliation{Department of Physics, Faculty of Sciences, King Abdulaziz University, Jeddah, Saudi Arabia}

\author{Kristina Todorovi\'c}
\affiliation{Department of Physics and Mathematics, Faculty of Pharmacy, University of Belgrade,
Vojvode Stepe 450, Belgrade, Serbia}

\author{Srdjan Kosti\'c}
\affiliation{Institute for the Development of Water Resources "Jaroslav \v{C}erni",\\
Jaroslava \v{C}ernog 80, 11226 Belgrade, Serbia}

\author{Nikola Buri\'c}
\email{buric@ipb.ac.rs}
\thanks{corresponding author}
\affiliation{Scientific Computing Laboratory, Institute of Physics, University of Beograd, P.O. Box 68, 11080 Beograd-Zemun, Serbia}

\begin{abstract}
We study the activation process in large assemblies of type II excitable units whose dynamics is influenced by two independent noise terms. The mean-field approach is applied to explicitly demonstrate that the assembly of excitable units can itself exhibit macroscopic excitable behavior. In order to facilitate the comparison between the excitable dynamics of a single unit and an assembly, we introduce three distinct formulations of the assembly activation event. Each formulation treats different aspects of the relevant phenomena, including the threshold-like behavior and the role of coherence of individual spikes. Statistical properties of the assembly activation process, such as the mean time-to-first pulse and the associated coefficient of variation, are found to be qualitatively analogous for all three formulations, as well as to resemble the results for a single unit. These analogies are shown to derive from the fact that global variables undergo a stochastic bifurcation from the stochastically stable fixed point to continuous oscillations. Local activation processes are analyzed in the light of the competition between the noise-led and the relaxation-driven dynamics. We also briefly report on a system-size anti-resonant effect displayed by the mean time-to-first pulse.
\end{abstract}


\maketitle

\section{Introduction}

For the wealth of local and intriguing collective phenomena displayed, the large assemblies comprised of excitable units have now been appreciated as a distinct class of dynamical systems. In terms of theory, the fundamental issue for understanding these systems is whether their macroscopic dynamics itself exhibits excitable behavior. The other important issue naturally concerns the relation between such macroscopic excitability and noise. So far, stochastic effects have been identified as a major factor contributing the collective behavior of systems of excitable units \cite{LGNS04,I07,KSMO10,ZNFS03,ZSSN05,TSTC07,FTVB12,GSRQ13,LXQZHM11,NM10,SSK07,QHHL10}. 

In this paper, we study large assemblies consisting of noisy type II excitable elements, which are represented by the canonical Fitzhugh-Nagumo ($FHN$) model. Conceptually, our focus will lie with two main points: $(i)$ demonstrating that an assembly made up of excitable units can itself be considered a macroscopic excitable element, and $(ii)$ identifying the analogies and pointing out the differences between the excitable behaviors of a single unit and an assembly.

Note that point $(i)$, at least to our knowledge, has not been treated explicitly so far. In particular, the main obstacle for analytically approaching this issue is that the macroscopic dynamics of an assembly of stochastic $FHN$ units cannot be expressed in a closed form via the global variables, which would otherwise make up a standard and the desired form of describing the collective motion. Some alternative forms of analysis are not available due to complexity of the corresponding Fokker-Planck equation that assumes an integro-differential form. In order to resolve this and explicitly demonstrate the excitability feature at the assembly level, we use an approach that relies on the mean-field ($MF$) approximate model of the assembly's collective motion. In several recent papers, the $MF$ model has already proven successful in the analysis of systems described by large sets of stochastic (delay) differential equations, in particular when treating stability of the stationary state, as well as the scenarios for the onset and the suppression of the collective mode \cite{BRTV10,FTVB13,FTVB14}. 

Analysis on point $(ii)$ can be carried out at two levels, one focused on the phenomenology involved, and the other concerning the statistical properties of the corresponding noise-driven activation processes. In terms of phenomenology, the excitability feature refers to capability of systems to generate spiking (pulse-like) responses or small-amplitude excitations, which are separated by some form of threshold. For a single unit, the large-amplitude response is comprised of the activation and the relaxation stage, such that the former is strongly influenced by noise, whereas the latter is typically deterministic and maintains a stereotype profile in a broad range of noise values. Among else, the stereotype character of pulse implies that its amplitude and width are independent on the form of perturbation applied.

The stated arguments on the notion of excitability should naturally hold in case of an assembly as well. Nevertheless, what may be different are the details related to the assembly's threshold-like behavior, which by itself stands out as a highly non-trivial issue. Another point of difference concerns the local mechanisms by which excitations of units within a population are elicited. In particular, each unit may be evoked to emit a pulse either by noise or via the interaction terms. Adhering to formulation of activation event for an excitable unit stated in our previous paper \cite{FTPVB14}, the latter point does not merely imply that the activation processes of individual units cannot be considered independent, as if they were driven by uncorrelated random perturbations, but more significantly indicates that the activation process of an arbitrary unit may be influenced by the relaxation processes of other units. The third point one should stress concerns the process of spike emission at the assembly level. In particular, having accepted the description of collective motion in terms of global variables, one should also bear in mind that their amplitude is affected by coherence of individual spikes, such that the increased coherence implies larger amplitude of the relevant macroscopic variable. By such a concept, evoking a large-amplitude excitation requires that the spikes for a sufficient fraction of individual units become coherent. Analyzing coaction of activation and relaxation processes for individual elements, as well as the role of spike coherence with respect to assembly pulse emission, present intricate issues, absent if small groups of units are considered.

Apart from phenomenological aspects, the comparative analysis of excitable behaviors of a unit and an assembly will address the respective noise-driven activation processes. In analogy to our paper on a single and two interacting excitable units \cite{FTPVB14}, we associate the term activation solely to the assembly's large-amplitude excitation. In principle, the definition of assembly activation problem should incorporate appropriate boundary conditions, which provide a clear-cut distinction between the small- and the large-amplitude excitations. Nevertheless, while a single most adequate and relevant definition for the activation problem may be given in cases of an excitable unit or a pair of units \cite{FTPVB14}, the specific character of global variables, or rather the fact that their amplitudes depend on coherence of individual spikes, prevents us from establishing such a formulation in case of an assembly of excitable elements. Instead, we introduce three alternative formulations of the assembly activation event which emphasize different aspects of the relevant phenomena. Their merit will depend on the aims of the particular study
of assembly dynamics and the fashion in which one can adapt them to potential applications.

Two of the formulations rest on the standard description of collective motion in terms of global variables, and are intended precisely at examining the analogy between the excitable behaviors of a single unit and an assembly. In particular, one formulation is consistent the threshold boundary approach for a $FHN$ unit (terminating boundary condition given by a threshold $x$-value), whereas the other derives from characteristic boundary approach (terminating boundary conditions given by an appropriate boundary set). Nevertheless, for comparison we also adopt a formulation where the assembly activation is treated as a compound event, comprised of \emph{only} first-pulse responses of a sufficient fraction of participant units, regardless of whether the emitted spikes are coherent or not. The implications of the three formulations are analyzed in terms of statistical properties of the corresponding activation events, characterized by the dependencies of the time-to-first pulse emission ($TFP$) and the related coefficient of variation on noise. 

The paper is organized as follows. In Sec. \ref{Model} is provided the background on the applied model. Section \ref{Excit} addresses the details of the assembly's excitable behavior, including the analysis carried out on the deterministic $MF$ approximation. In Section \ref{Defs} are laid out the three formulations of the assembly activation event. Section \ref{Stats} contains the detailed numerical analysis on the statistical properties of activation events conforming to the three adopted formulations. We also consider the qualitative explanation for the bimodal or unimodal distributions of local activation events typical for certain domains of noise intensities. Subsection \ref{Size} concerns the effects related to system size, including the "anti-resonance" found for the mean $TFP$s at fixed noise intensities. In Sec. \ref{Conc} is given a brief summary of our main points.

\section{Details of the applied model}
\label{Model}

As a paradigm for analyzing collective excitable behavior, we consider an assembly comprised of $FHN$ units.
The dynamics of an arbitrary unit $i$ is given by
\begin{align}
dx_i&=(x_i-x_i^3/3-y_i)dt+\frac{c}{N}\sum_{j=1}^N (x_j-x_i)dt+\sqrt{2D_1}dW_1^i \nonumber\\
dy_i&=\epsilon(x_i+b)dt+\sqrt{2D_2}dW_2^i, i=1,\dots N \label{eq1}
\end{align}
where $b$ and $\epsilon$ are the intrinsic unit parameters, while $c$ denotes the coupling strength. The units
are assumed to be identical and are connected in the all-to-all fashion. Parameter $\epsilon$ is set to a small
value $\epsilon=0.05$, which warrants that the characteristic time scales for $x_i$ and $y_i$ evolution are sharply separated. Being type II excitable means that the units are poised close to transition toward oscillatory state
via the Hopf bifurcation \cite{I07}. The bifurcation parameter $b$ is set to $1.05$, the value just below critical threshold. Excitability feature of a single $FHN$ unit has been extensively analyzed \cite{LGNS04,I07,KPLM13}, and an overview can also be found in our preceding paper \cite{FTPVB14}. At variance with the latter, the perturbation here may either arrive from the interaction terms, or may be caused by random fluctuations due to two independent sources of noise. Motivated by the possible interpretation in the field of neuroscience \cite{DRL12,KPLM13}, we adopt the convention by which the stochastic terms in the fast (slow) variables are referred to as external noise $D_1$ (internal noise $D_2$). Note that $dW_k^i, k\in\{1,2\},i=1,\dots,N$ denote stochastic increments of independent Wiener processes whose averages and correlations satisfy $\langle dW_k^i\rangle=0$, $\langle dW_k^idW_l^j\rangle=dt\delta_{kl}\delta{ij}$.

In order to gain insight into the assembly's collective dynamics, one may first carry out the bifurcation
analysis of the deterministic (noiseless) version of system \eqref{eq1}. For $N$ sufficiently large so that
the terms $\mathcal{O}((c/N)^2)$ and of higher order can be neglected, the characteristic equation describing the stability of equilibrium $(x_1,y_1,x_2,y_2,\dots,x_N,y_N)=(-b,-b+b^3/3,-b,-b+b^3/3,\dots,-b,-b+b^3/3)$ is given by an approximate expression
\begin{equation}
(\lambda^2-(1-b^2)\lambda+\epsilon)(\lambda^2-(1-b^2-c)\lambda+\epsilon)^{N-1}=0. \label{eq2}
\end{equation}
Since $b=1.05$ is kept fixed, it follows that the equilibrium may become unstable only via the direct supercritical Hopf bifurcation controlled by $c$. Nonetheless, the critical $c$ value is $c^*=1-b^2<0$, which implies that the positive coupling strengths do not affect the stability of equilibrium. In the present paper, we only consider the subcritical values $c>0$, such that the system \eqref{eq1} always lies in the excitable regime.

As for the impact of stochastic fluctuations on the asymptotic collective dynamics, it is known that the noise intensity may act as a control parameter in excitable media, giving rise to three generic regimes of macroscopic behavior \cite{KSMO10}. In particular, when noise is systematically increased, the dynamics of global variables undergoes a sequence of transitions, first exhibiting the stochastically stable equilibrium, then the stochastically stable limit cycle, and eventually it decays into disordered behavior. This sequence may be explained as follows. In the approximately stationary state, at any given moment, most of the units lie in the vicinity of equilibrium, whereas the relatively rare excursions due to weak noise or the interaction terms remain incoherent. Therefore,
the macroscopic variables are only marginally different from equilibrium values of single elements. At some point, the increase of noise induces more or less coherent oscillations of units which are easy to synchronize. This conforms to the onset of the collective mode according to the scenario of stochastic bifurcation \cite{A99,GLV09,GBV11,ABR04}. In the supercritical state, the global variables follow a limit cycle attractor whose profile is similar to that of relaxation oscillations of individual units. Once the noise intensity becomes strong enough to overcome the effect of couplings, the disordered regime sets in. The spiking frequency of units remains high, but their activity desynchronizes. Since the majority of units at any moment are refractory, global variables display irregular oscillations with a quenched amplitude.

From our perspective, the conceptually most important transition is the one associated to occurrence of noise induced collective oscillations. This phenomenon is relevant for the activation process because it indicates the loss of stochastic stability for the fixed point. In other words, under increasing noise, the attractive ability of the fixed point gradually reduces and is eventually lost, which naturally affects how the phase point escapes the vicinity of equilibrium. Note that the above sequence of states has previously been verified in case of an assembly of $FHN$ elements driven by internal noise. We have found that the similar sequence persists if the units are influenced by external noise, though the individual activities become more difficult to synchronize, rendering the amplitude of collective oscillations comparably smaller than the one emerging under internal noise.

\section{Excitable behavior at the assembly level}
\label{Excit}

Having summarized the points relevant for the deterministic part of dynamics described by \eqref{eq1}, we turn to characterization of the assembly's excitability feature. The standard approach to collective motion is to introduce the macroscopic variables $X(t)=\langle x_i(t)\rangle=\frac{1}{N}\sum\limits_{i=1}^Nx_i(t)$ and
$Y(t)=\langle y_i(t)\rangle=\frac{1}{N}\sum\limits_{i=1}^Ny_i(t)$, whereby the aim typically lies in establishing some form of analogy between the dynamics of single units and an assembly. The latter would be especially relevant
for examining the issue of excitability at the assembly level. Nevertheless, it is evident that the compound effect of nonlinear terms prevents the whole assembly to evolve in the fashion analogous to that of a single unit, which would occur only if $\langle x_i^3\rangle=\langle x_i\rangle^3$ were to hold. Therefore, the collective dynamics in principle cannot be expressed in a closed form via the global variables. The other potential approaches to analysis of macroscopic excitable behavior are severely limited by the difficulties associated to Fokker-Plank formalism, where the equation for the one-particle density $P(x,y,t)$ acquires a complex integro-differential form $\frac{\partial}{\partial t}P=-\frac{\partial}{\partial x}[x(1-c)-\frac{x^3}{3}-y+c\int x_1P(x_1,y_1,t)dx_1dy_1]P-\frac{\partial}{\partial y}\epsilon(x+b)P+D_1\frac{\partial^2P}{\partial x^2}+D_2\frac{\partial^2P}{\partial y^2}$.

The arguments above imply that one is required to introduce approximations for collective dynamics
corresponding to the \emph{deterministic} part of system \eqref{eq1} in order to analyze the assembly
excitability feature and the associated threshold-like behavior. In the following, we consider two
approximate models of collective motion, distinguished by the fashion in which the effect of interaction
terms is resolved. The first model holds if the interaction terms vanish or can be neglected. Note that this condition is satisfied if the initial conditions for all the units are identical or lie close to each other.
Since the evolution of the system is deterministic and the coupling is diffusive, such selection of initial conditions facilitates that the whole assembly acts as a macroscopic excitable element. The evolution of global variables is then given by the equations analogous to those for a single unit
\begin{align}
dX&=(X-X^3/3-Y)dt\nonumber\\
dY&=\epsilon(X+b)dt.\label{eq25}
\end{align}
The details regarding the excitability feature of such a system are well established \cite{KPLM13}. In particular,
recall that its threshold behavior is associated to "ghost-separatrix", a thin layer made up of canard-like
trajectories that foliate around the maximum of the fast-variable nullcline, whereby the spread increases with
the characteristic scale separation ratio $\epsilon$.

Let us now consider a more sophisticated approximation that takes into account the net effect of the interaction terms. The analytical framework suitable for demonstrating the assembly excitability feature in this more general case is provided by the mean-field approach. Before laying out the details, we make an overview of the ingredients crucial for the derivation of the $MF$ model, as well as the results achieved so far on treating the systems of large sets of stochastic (delay)-differential equations. In principle, deriving the $MF$ model involves a number of nontrivial elements, and it ultimately leads to a deterministic system amenable to standard bifurcation analysis, where noise intensity may act as a bifurcation parameter. The $MF$ method combines the cumulant approach with Gaussian approximation, according to which all the cumulants above second order are assumed to vanish. The latter is intended as a closure hypothesis, which is necessary due to presence of nonlinear terms in the exact system \eqref{eq1}. Thus, starting from the original system which in general comprises $kN$ (delay) differential equations, where $k$ is the number of local degrees of freedom, one ends up with a set of $k(k+3)/2$ deterministic (delayed) equations describing the evolution of the means, as well as the appropriate variances and the covariances.

As for the main results achieved thus far, the $MF$ method has already been applied in analyzing the stability of assemblies of (delay)-coupled excitable elements, as well as the scenarios for the onset and the suppression of the collective mode. In particular, the bifurcations displayed by the $MF$ model can qualitatively account for the stochastic bifurcations which the exact system undergoes \cite{LGNS04,ZNFS03,ZSSN05,BRTV10,FTVB13,SPRKSG14,SSG13,SZNSG13}. It has also been shown that the $MF$ model can provide accurate quantitative predictions, reflected in a close agreement between the oscillation period of the $MF$ model and the average oscillation period of the exact system \cite {BRTV10,FTVB13}. Furthermore, the $MF$ approach has proven successful in the analysis on stability and the onset/suppression of the collective mode in case of interacting assemblies, thereby indicating a potential extension to modular networks \cite{FTVB13}. We stress that the approximations behind the $MF$ model used here, called the Gaussian approximation and the quasi-independence approximation, have been analyzed in detail, not only in terms of precise formulations and adaptation for the systems of class II excitable units, but also with respect to parameter domains that warrant their validity \cite{FTVB14}.  In this context, an important finding is that the dynamics of the $MF$ model can indicate in a self-consistent fashion the parameter domains where the $MF$ approximations break down.

The $MF$ model corresponding to \eqref{eq1} involves five equations describing the evolution of the means $m_x(t)=E(x_i(t)), m_y(t)=E(y_i(t))$, the variances $s_x(t)=E((x_i(t)-\langle x_i(t)\rangle)^2), s_y(t)=E((y_i(t)-\langle y_i(t)\rangle)^2)$ and the covariance $u(t)=E((x_i(t)-\langle x_i(t)\rangle)( y_i(t)-\langle y_i(t)\rangle))$. Note that $E(\cdot)$ denotes expectations over an ensemble of different stochastic realizations. The detailed derivation of the $MF$ model may be found in \cite{BRTV10,FTVB13}, whereas here we just state
the final result
\begin{align}
{\dot{m_x}}&=m_x-m_x(t)^3/3-s_xm_x-m_y,\nonumber\\
{\dot{m_y}}&=\epsilon(m_x+b),\nonumber\\
{\dot{s_x}}&=2s_x(1-m_x^2-s_x-c)-2u+2D_1\nonumber\\
{\dot{s_y}}&=2\epsilon u+2D_2,\nonumber\\
{\dot{u}}&=u(1-m_x^2-s_x-c)+\epsilon s_x-s_y.\label{eq3}
\end{align}
As already indicated, the influence of stochastic terms is expressed through the noise intensities $D_1$ and $D_2$.

Nevertheless, in order to make the analogies between the excitable behaviors of a single $FHN$ unit and an assembly explicit, one should arrive at the system describing the collective motion by two equations. The latter would allow one to apply the phase plane analysis derived from the framework of singular perturbation theory. This is achieved by introducing an additional "adiabatic" approximation \cite{LGNS04}, which consists in assuming that the relaxation of second-order moments is much faster than that of the first-order ones. This is not a crude approximation given that the initial conditions of units in the exact system are set to be identical (coinciding with the deterministic fixed point), especially if the noise intensities are not too large. Having replaced the fast variables with the stationary values, one obtains the following system for the dynamics of the means:
\begin{align}
\frac{dm_x}{dt}&=F(m_x,m_y)=m_x-\frac{1}{3}m_x^3-m_y -\frac{m_{x}}{2}(1-c-\nonumber \\
&m_{x}^2+\sqrt{(1-c-m_{x}^2)^2+4(D_1+D_2/\epsilon}) \nonumber \\
\frac{dm_y}{dt}&=G(m_x,m_y)=\epsilon(m_x+b).\label{eq4}
\end{align}
An apparent advantage of \eqref{eq4} is that one may extend the results of phase plane analysis to assembly dynamics, thereby gaining insight into whether and how the excitability feature is manifested at the level of global variables.

\begin{figure}
\centerline{\epsfig{file=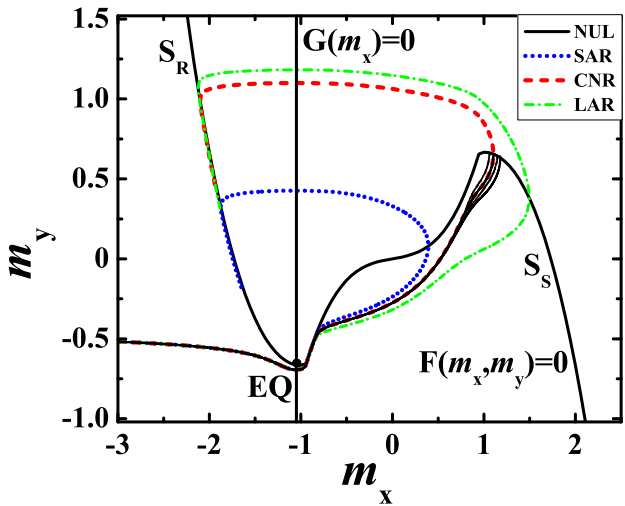,width=8.5cm}}
\caption{(Color online) Characterization of the assembly excitable behavior via the phase plane analysis of the corresponding $MF$ model \eqref{eq4}. The equilibrium ($EQ$) lies at the intersection of the nullclines $F(m_x,m_y)=0$ and $G(m_x)=0$. The $m_x$-nullcline is comprised of three branches, whereby the spiking and the refractory branches $S_S$ and $S_R$ are attractive. For finite $\epsilon$, the boundary between the sets of initial conditions that lead to small- or large-amplitude responses, $SAR$ and $LAR$ respectively, foliates into a thin layer of canard-like trajectories ($CNR$), which we refer to as "ghost separatrix" (solid black lines). The trajectories belonging to the boundary layer are obtained by fixing the $m_x$ initial condition to a particular value $m_{x,0}=-3$, while sweeping over $m_{y,0}$. The fact that a difference in $m_{y,0}$ of the order of $10^{-18}$ evokes a different type of response corroborates extreme sensitivity to initial conditions in vicinity of ghost separatrix. The results shown refer to case $\epsilon=0.07,b=1.05$.}
\label{Fig1}
\end{figure}

The main point is that the system \eqref{eq4} displays class II excitable behavior, which provides an indication that the collection of excitable $FHN$ units described by \eqref{eq1} itself constitutes a macroscopic excitable system. The $m_x$ and $m_y$ nullclines, as well as an illustration of how the two marginally different initial conditions give rise to small- or large-amplitude responses are provided in Fig. \ref{Fig1}. Note that the results are obtained for system \eqref{eq4} under $D_1=D_2=0$. The $m_x$-nullcline again consists of three branches, such that in the singular limit $\epsilon\rightarrow0$ the spiking branch $S_S$ and the refractory branch $S_R$ are attractive, whereas the middle branch is unstable. Compared to the case of single unit, the profile of the middle branch is changed and includes a flexion point, which may be attributed to the compound effect of interaction between the units. The two types of population response to perturbation, or rather the associated threshold-like behavior, imply the existence of a soft boundary between the corresponding initial conditions.

Carrying out the analysis analogous to that for a single unit, described in brief in the caption of Fig. \ref{Fig1}, one may show that the boundary is again given by a thin layer of system trajectories which foliate around the maximum of the $m_x$-nullcline. The relevant segments of such trajectories are indicated in Fig. \ref{Fig1} by the black solid lines. What is found is quite reminiscent to "ghost-separatrix" in case of a single excitable unit \cite{KPLM13}. The foliation here also becomes more pronounced with increasing $\epsilon$. Compared to those for a single unit \cite{KPLM13}, the trajectories that make up the boundary layer are seen to converge further away from the maximum, mainly because the unstable branch in case of an assembly involves a flexion point. Also, the numerical evidence suggests that the mean-field variables, and therefore the collective dynamics of the assembly, shows greater sensitivity to initial conditions compared to that of a single unit. This conclusion follows from the fact that under analogous parameter values, obtaining the trajectories that belong to boundary set requires a higher numerical precision in case of the $MF$ model than for a single unit, cf. \cite{FTPVB14} and Fig. \ref{Fig1}.

Note that in Sec. \ref{Stats} we use the $MF$ model to gain insight into the stochastic stability of equilibrium of the exact system, or rather the latter's sensitivity to random perturbations. This point will be crucial for explaining how the form of the $\tau(D_1,D_2)$ dependence is related to the stochastic bifurcation leading from the stochastically stable fixed point to continuous oscillations.

\section{Alternative formulations of assembly activation problem}
\label{Defs}

As announced in the Introduction, we consider three alternative formulations of the activation problem, which are associated to different definitions of the assembly activation event. Since the conceptual aspects of a large assembly's activation problem have not been treated so far, one cannot \emph{a priori} hold any formulation preferred over the others. Thus, the implications of each of the formulations will be qualitatively analyzed and then compared. Note that the physical picture laid out here is distinct from the cases of a single and two coupled excitable units treated in our previous paper \cite{FTPVB14}, where we have been able to provide a unique adequate formulation of the activation event, which among else, allows an immediate generalization from a single unit to a two-unit setup. To properly address the issue of assembly activation, one should invoke a couple of remarks from that study.

First, the activation problem we consider is associated to pulse emission, and as such cannot be viewed as an extension of a typical escape problem. Note that the latter would require a genuine saddle structure at the terminating boundary \cite{BMLSM05}, which implies coexistence between two attractors, whereas in our problem
the fixed point is the only relevant, and often the unique attractor. In terms of specifying the terminating boundary, an important point has been to replace the somewhat arbitrary notion of "threshold" for pulse emission by the relevant boundary set consistent with the underlying structure of the phase space. Finally, in case of two units, we have argued that the definition of activation event where the phase points of each unit are supposed to reach the appropriate terminating boundary is preferred over any formulation involving the two-unit averages $(x_1(t)+x_2(t))/2$ and $(y_1(t)+y_2(t))/2$. This applies because the averages contain additional information on synchronization of units, which are secondary to the two-unit activation process. Nevertheless, such an approach can only be maintained for small groups of units. For larger assemblies, it is of interest to state the activation problem in terms of global variables, since they present a standard tool for describing collective motion, both theoretically and in applications.

The point which makes the case of large populations intriguing concerns the underlying mechanisms of activation.
In particular, the processes on a local and global level are not influenced only by noise, but are also strongly contributed by the relaxation of units. The arguments above further suggest that an elaborate study on analogies and differences between the statistical properties of activation process in small groups of units and the large assemblies would be in order. Therefore, our approach is on one hand to retain the formulation of the assembly activation problem inherited from the two-unit case where the global variables are not considered, and on the other hand, to introduce two additional formulations that explicitly refer to dynamics of global variables. These three formulations are specified as follows.

{\textbf{Formulation $1$.}{\it The assembly activation event occurs when more than a half of participant units have emitted their first pulses.} According to this, assembly activation is perceived as a compound event made up of local activation events. For the local events we adopt the definition provided in our previous paper \cite{FTPVB14}: the activation path of a single excitable unit influenced by noise emanates from the deterministic fixed point and terminates at the boundary set coinciding with the spiking branch of the limit cycle which would exist in the corresponding supercritical state. In the present context, the terminating boundary set can with sufficient accuracy be approximated by the spiking branch of the cubic nullcline for a non-interacting unit. Motivation behind formulation $1$ draws in part from certain applications, especially in the field of neuroscience, where population response to external stimuli typically engages a certain fraction of units, rather than the entire assembly \cite{BP12,HDZ08,KKGBSH07,VG04,A03,BH99}. From the qualitative perspective, demanding any reasonable macroscopic fraction other than a half of units to be activated makes as good a choice as any, because the main statistical properties of the ensuing activation process will remain similar. In a sense, formulation $1$ can be interpreted as an extension of the definition introduced for a two-unit activation event in our previous paper \cite{FTPVB14}.

{\textbf{Formulation $2$.}{\it The assembly activation event occurs if the global variable $X(t)$ crosses the predefined threshold $X_0$}($X(t)>X_0,X'(t)>0$). Unlike the case of a single unit \cite{KPLM13,FTPVB14}, the formulation involving an explicit threshold is justified for the global variable because its amplitude depends on coherence of spikes of single units. The latter point introduces ambiguities when attempting to analyze the assembly threshold-like behavior. In other words, for a single unit, it is not difficult to distinguish between the small and large amplitude responses, given that the amplitude of superthreshold excitations is stereotypical. However, in case of an assembly, one is able to understand the associated threshold-like behavior only in terms of the approximate $MF$ model, cf. Sec. \ref{Excit}, but cannot provide clear-cut criteria in terms of global variables of the exact system, especially if coherence of units' activities for the given parameter set is weak. This point will be further explained when discussing formulation $3$, while here we just mention that noise domains may be found where formulation $2$ is more or less suitable. Compared to formulation $1$, formulation $2$ is conceptually distinct because the assembly activation events can be affected by multiple spikes of individual units. The potential consequences of selecting particular threshold values $X_0$ will be discussed in the next Section.

{\textbf{Formulation $3$.}{\it The assembly activation event occurs once the phase point associated to global variables reaches the appropriate terminating boundary set, defined in analogy to a single unit case.} This formulation involves an implicit assumption that the threshold-like behavior of an assembly is qualitatively similar to that of individual units. As indicated in Sec. \ref{Excit}, if the condition $\langle x_i^3(t)\rangle\approx \langle x_i(t)\rangle^3$ is fulfilled, the dynamics of global variables can be approximated by the equations analogous to that of a single unit. Consistent with this approximation, the collective motion of finite assemblies
in presence of noise may be described by \cite{KSMO10,TMG03}
\begin{align}
dX&=(X-X^3/3-Y)dt+\sqrt{\frac{2D_1}{N}}dW_1\nonumber\\
dY&=\epsilon(X+b)dt+\sqrt{\frac{2D_2}{N}}dW_2.\label{eq45}
\end{align}
As a corollary, one may use the phase plane analysis and describe the assembly first-pulse emission in terms of motion of phase point $(X,Y)$ to the spiking branch of the corresponding $X$-nullcline, which is then naturally considered the relevant terminating boundary set for the assembly activation problem. Nevertheless, the given physical picture is valid only if the fluctuations of individual units are small enough for the above condition to apply, which ultimately depends on the noise intensities.

\begin{figure}
\centerline{\epsfig{file=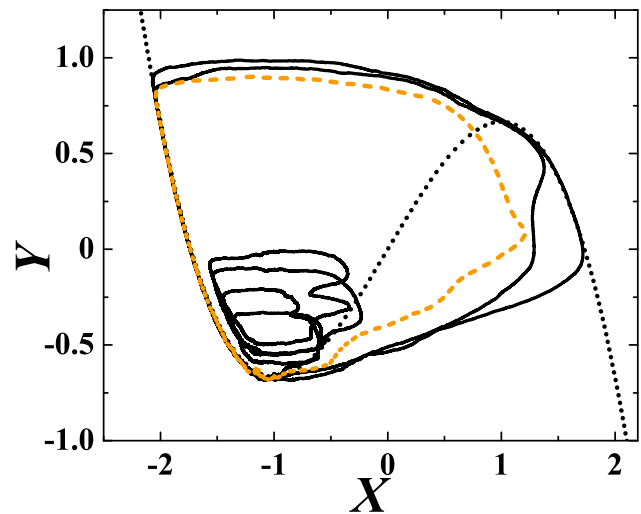,width=8.5cm}}
\caption{(Color online) Illustration of differences between formulations $2$ and $3$. For certain $(D_1,D_2)$, the assembly threshold-like behavior is difficult to resolve because the amplitude of large fluctuations is not stereotypical. The presented $(X,Y)$ orbit is obtained for a \emph{single} stochastic realization under parameter
set $D_1=0.0004,D_2=0.0008,c=0.1,N=50$. The bold dotted line indicates the $X$-nullcline consistent with the approximate system \eqref{eq45}. Apart from small-amplitude excitations that remain in close vicinity of deterministic fixed point, one also finds substantially larger excitations far above the stochastic fluctuations around the "stable state", where the phase point may either fall short of the spiking branch of the $X$-nullcline (segment of trajectory shown by the orange dashed line), or may actually reach it (segment indicated by the solid black line). The former instance conforms to activation event by formulation $2$, but does not comply with formulation $3$. The criteria as to which large excitation is considered an assembly activation ultimately depends on the scope of the particular study.}
\label{Fig2}
\end{figure}

\begin{figure*}
\centerline{\epsfig{file=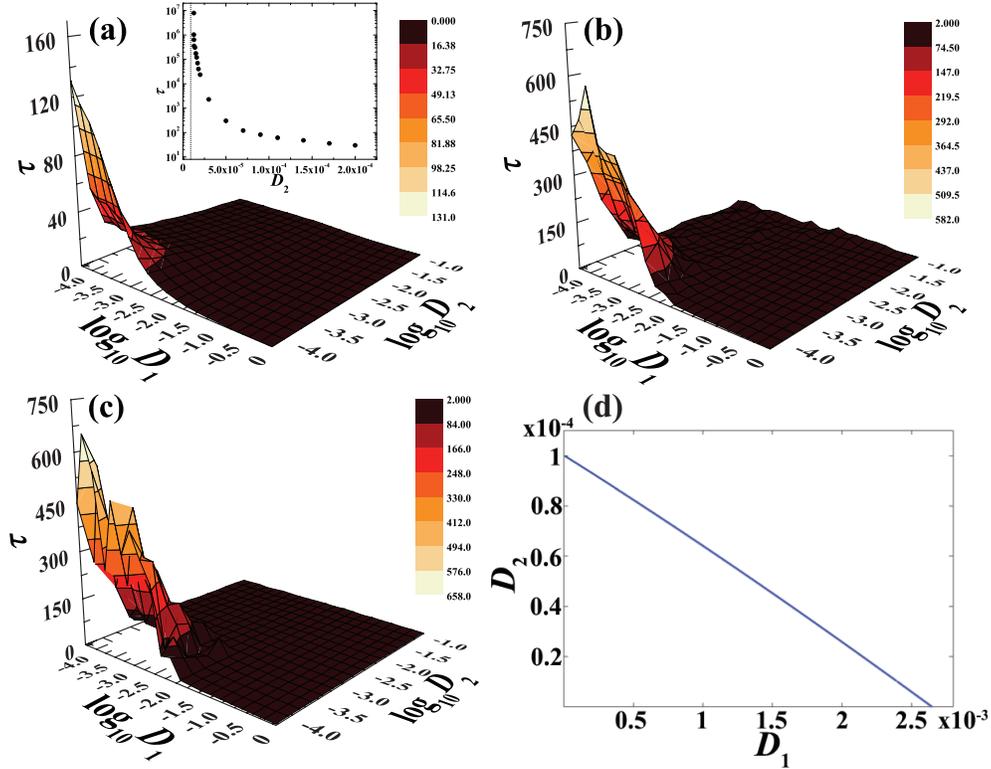,width=13cm}}
\caption{(Color online) Mean $TFP$s for the assembly activation process and the role of stochastic bifurcation.
(a), (b) and (c) show $\tau(D_1,D_2)$ dependencies for formulations $1-3$ of the activation event, respectively. Results are obtained for $c=0.1$ and the assembly size $N=100$, whereby the stochastic average is taken over an ensemble of $300$ different activation paths. The inset in (a) illustrates exponential divergence of $\tau(D_2)$ for fixed $D_1=10^{-5}$ as one approaches $D_2^*\approx1.25*10^{-5}$ (the value indicated by the dotted line). The qualitative similarity between the $\tau(D_1,D_2)$ dependencies in (a), (b) and (c) is associated to stochastic bifurcation underlying transition to continuous oscillations. The corresponding bifurcation curve $D_2(D_1)$, determined by analyzing $MF$ model \eqref{eq4}, is shown in (d).}
\label{Fig3}
\end{figure*}

An important note on formulation $3$ is that \emph{coherence} between spikes (approximate matching of spike times) of a large fraction of individual units is required to elicit an assembly activation event, viz. the first-pulse emission for the global variables. For this point, let us make additional remarks on the issues of terminating boundary conditions and the role of spike coherence in the assembly activation process. First, there is no principle difference in the dynamics underlying activation scenarios for three formulations of the assembly activation event. The differences in statistical features, which will be discussed in the next Section, are caused by the selection of terminating boundary conditions. This is made explicit in Fig. \ref{Fig2}, where a \emph{single} stochastic orbit for $(X,Y)$ is used to illustrate the difference between formulations $2$ and $3$. According to formulation $2$, the assembly has emitted a pulse once $X$ crosses a certain threshold (segment of the trajectory indicated by the dashed red line), while formulation $3$ requires that the phase point reaches the spiking branch of the $X$-nullcline (segment of the trajectory indicated by the solid black line). Thus, considering the very same stochastic trajectory for $(X,Y)$, the assembly activation event conforming to formulation $2$ would have happened earlier than the one satisfying formulation $3$. Nevertheless, both of the considered sections of the $(X,Y)$ orbit correspond to large excitations far above the stochastic fluctuations in vicinity of equilibrium. In fact, which excitation should be considered an activation event effectively depends on the scope of the study. In this context, formulation $2$ is more adapted to practical applications, whereas formulation $3$ has a more theoretical background.

Comparing Fig. \ref{Fig1} and Fig. \ref{Fig2}, one should point out that the latter captures certain fine details on collective dynamics that cannot be reflected in the former. This is because the physical background of Fig. \ref{Fig1} lies in the approximate $MF$ model, where the stochastic fluctuation effects above the second order are averaged out. In other words, the behavior of the exact system in Fig. \ref{Fig2} is associated with large fluctuations, which can no longer be described by the $MF$ model.

\section{Statistical properties of activation process}
\label{Stats}

\subsection{Mean $TFP$s and $TFP$ variability}
\label{means}

The statistical properties of the assembly activation process influenced by external and internal local noise are characterized by the mean $TFP$ $\tau(D_1,D_2)$ and the associated coefficient of variation $R(D_1,D_2)$ \cite{PPS05,PPM05,HE05}. Both quantities involve averaging over an ensemble of different stochastic realizations. In particular, the mean $TFP$ is given by $\tau(D_1,D_2)=\langle\tau_k\rangle=\frac{1}{n_r}\sum\limits_{k=1}^{n_r}\tau_k(D_1,D_2)$, where $n_r$ denotes the number of realizations, while the expression for the coefficient of variation reads $R(D_1,D_2)=\frac{\sqrt{\langle \tau_k^2\rangle-\langle \tau_k\rangle^2}}{\langle \tau_k\rangle}$. Since $R$ is variation of $TFP$s normalized by the mean, its smaller values indicate a better clustering of individual $TFP$s around $\tau$ \cite{PK97}. Note that the numerical simulations are carried out via Heun integration scheme with the fixed time step $\delta t=0.002$. The results for the mean $TFP$s and the variances are obtained by averaging over an ensemble of at least $300$ different stochastic realizations of the activation process. For each realization, the initial conditions for all units are identical and are given by the deterministic fixed point of system \eqref{eq1}.

The fields $\tau(D_1,D_2)$ corresponding to formulations $1-3$ of the assembly activation event are plotted in Figs. \ref{Fig3}(a), \ref{Fig3}(b) and \ref{Fig3}(c), respectively. Note that all three dependencies exhibit \emph{qualitatively} similar behavior, whereby one can clearly discern the domain of large mean $TFP$s and the plateau region. We find that these analogies derive from the fact that the profile of $\tau(D_1,D_2)$ dependencies in each of the instances is crucially influenced by the stochastic bifurcation. The latter corresponds to the noise-induced transition from the stochastically stable fixed point to continuous oscillations. It is intuitively clear that the stochastic bifurcation should be associated to boundary between the two characteristic forms of $\tau$ behavior, because above the bifurcation, the fixed point loses its attractive power, which makes it easier for noise to induce large-amplitude fluctuations. Also, once the attractive power of the fixed point is lost, any further increase of noise cannot induce qualitatively novel effects. This fact accounts for the existence of the large plateau region in the $\tau(D_1,D_2)$ dependence. As already indicated, we may gain insight into stochastic bifurcation by carrying out bifurcation analysis on the $MF$ counterpart \eqref{eq4} of the exact system. In particular, it is demonstrated that the $MF$ model undergoes direct supercritical Hopf bifurcation which qualitatively reflects the stochastic bifurcation of the exact system. The Hopf bifurcation curve $D_2(D_1)$ obtained for the $MF$ model at fixed $c=0.1$ is shown in Fig. \ref{Fig3}(d).

Note that for small $(D_1,D_2)$ the mean $TFP$s grow extremely large. This is explicitly illustrated in the inset of Fig. \ref{Fig3}(a), which shows shows that $\tau(D_2)$ (for fixed small $D_1$) exponentially diverges as a certain value of noise intensity $D_2^*$ is approached. The latter point clearly demonstrates the existence of potential barrier associated to the assembly's activation process. The lower boundary on noise intensities that facilitate activation is the largest for formulation $3$, and decreases for formulations $2$ and $1$. The reason behind such a behavior will be considered below.

\begin{figure}
\centerline{\epsfig{file=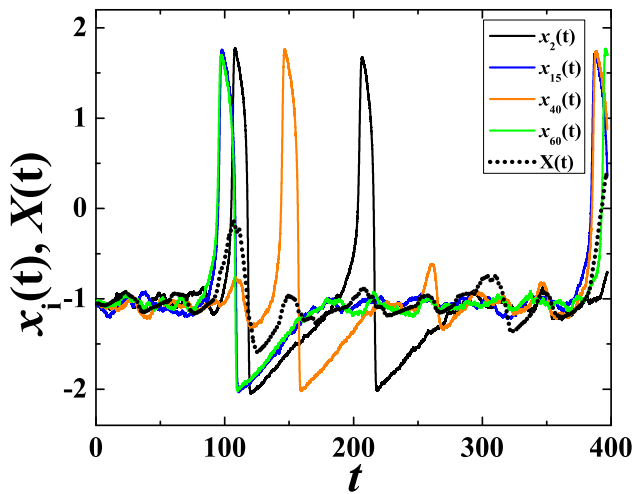,width=8.5cm}}
\caption{(Color online) Build-up of spike coherence prior to assembly activation event. The data concern a \emph{single} stochastic realization of the assembly activation event consistent with formulation $2$ ($X_0=0.4$). The dotted line shows the $X(t)$ dependence, whereas the solid lines refer to the appropriate time series of individual units $x_i(t)$. The system parameters are $D_1=0.0001365,D_2=0.0001,c=0.1,N=100$.}
\label{Fig9}
\end{figure}

In quantitative terms, the $\tau$ dependencies for formulations $1-3$ are manifestly different in
the regime subcritical relative to stochastic bifurcation. In particular, Fig. \ref{Fig3} shows that the corresponding $\tau$ values for the same noise intensities are typically ordered as $\tau_1<\tau_2<\tau_3$.
This may be explained by looking into the role played by the spike coherence between the individual units in the respective assembly activation processes. Let us focus first on the adjustment between the local time series $x_i(t)$ prior to assembly activation, see Fig. \ref{Fig9}. Naturally, the early spikes of single units will be the least coherent because the firing is driven by noise, and the effect of diffusive coupling in bringing their firing times closer is weakly felt. As the time passes, the impact of coupling on units' dynamics is felt more strongly, which gradually leads to a \emph{build-up of spike coherence} within the assembly. In other words, firing of individual units will become more coherent (more spikes fall within a relatively narrow time window) as the result of interactions. With time, such approximate synchronization may comprise an increasing number of units, depending on the $c$ value.

\begin{figure}
\centerline{\epsfig{file=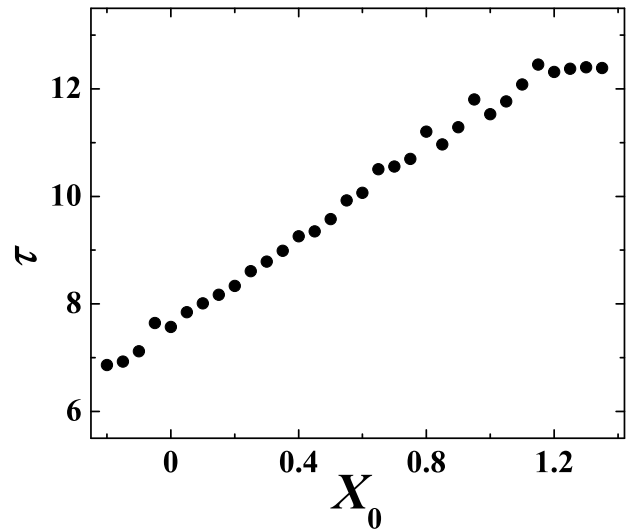,width=8.5cm}}
\caption{Impact of different $X_0$ on the results for formulation $2$ of activation event. The example
provided illustrates the form of $\tau(X_0)$ dependence typical in a broad range of $(D_1,D_2,c,N)$ values.
One finds similar behavior for all the other statistical quantities considered. The particular data are obtained
for $D_1=0.0207018,D_2=0.00101739,c=0.1,N=100$, with the stochastic averaging carried out over an ensemble
of $500$ different activation paths. The results imply that the statistical properties of activation process
derived from formulation $2$ are qualitatively independent on $X_0$.}
\label{Fig4}
\end{figure}

\begin{figure*}
\centerline{\epsfig{file=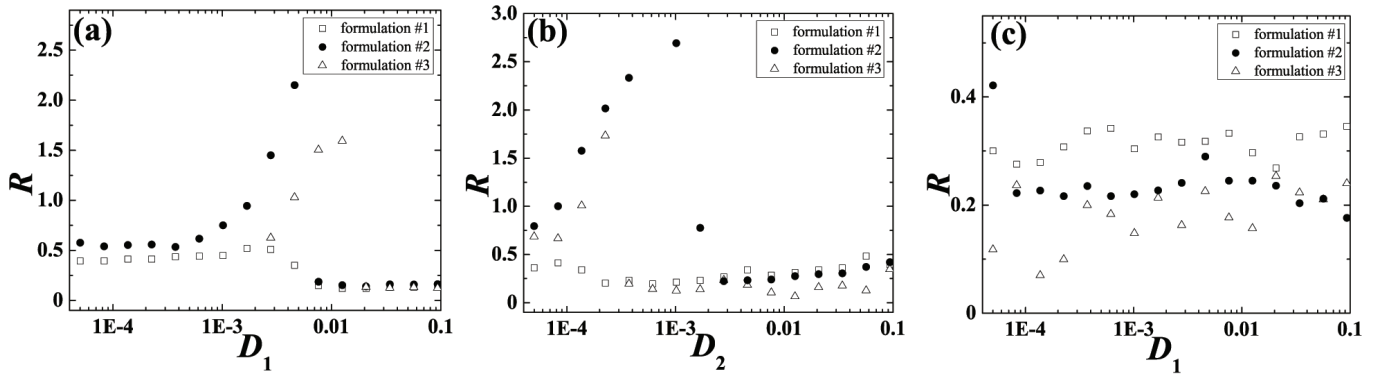,width=18cm}}
\caption{Impact of two noise sources on variability of assembly activation process. The $R$ values corresponding to formulations $1-3$ of the activation event are represented by open squares, solid circles and open triangles, respectively. In (a) is provided the $R(D_1)$ dependence for fixed $D_2=0.00005$, whereas (b) shows the $R(D_2)$ behavior at fixed $D_1=0.00062$. (c) refers to $R(D_1)$ dependence for $D_2=0.00459$. The noise values kept fixed
in (a) and (b) are small, such that the system undergoes stochastic bifurcation under variation of $D_1$ in (a) and $D_2$ in (b). The $D_2$ value in (c) is supercritical relative to stochastic bifurcation. The remaining system parameters are $c=0.1, N=100$.}
\label{Fig5}
\end{figure*}

Now we consider how this reflects to assembly activation for different formulations of the activation event. Formulation $1$ requires only that more than half of units have emitted their first spikes, imposing no demand for their coherence. In line with the above arguments, satisfying such a requirement will take the least amount of time, which means that the pertaining $\tau$ will be smaller compared to mean $TFP$s for formulations $2$ and $3$ under the same parameter set. Formulations $2$ and $3$ concern the global variables and thereby do require a certain level of spike coherence within the assembly. Formulation $3$ imposes a more strict condition, so its fulfillment is associated to the largest level of spike coherence. Achieving larger level of spike coherence takes more time, so in the subcritical regime $TFP$s for formulation $3$ will necessarily be larger than those for formulation $2$. It is also apparent that the assembly activation will necessarily be induced by the first spikes of single units only for formulation $1$, whereas activation by formulations $2$ and $3$ is bound to be contributed by latter spikes of individual units. Above the stochastic bifurcation, coherence of spikes between the units is naturally achieved, such that the assembly activation for all three formulations is contributed by the first spikes of units. This is the reason why the mean $TFP$s for $(D_1,D_2)$ above the bifurcation have similar values for all three formulations of the activation event.

Before proceeding, we make a brief remark on certain details related to formulation $2$ of the activation event. As indicated in Sec. \ref{Defs}, it has to be verified whether the results on statistical properties of the activation process depend on the particular value of the threshold $X_0$. We have established that for any reasonable selection, viz. the $X_0$ values lying sufficiently above the amplitude of stochastic fluctuations around the fixed point, the corresponding dependencies $\tau(D_1,D_2)$ maintain qualitatively similar profile. This holds for $X_0$ values within the range $X_0\in[-0.1,1.3]$. Considering $\tau$ values as an example, we have shown that $\tau(X_0)$ for arbitrary $(D_1,D_2)$ monotonously increases until reaching saturation around $X_0\approx1.2$, see Fig. \ref{Fig4}. The analogous conclusions regarding the qualitative independence of the results on $X_0$ have been confirmed for all the other statistical properties of the activation process.

Let us now examine the impact of two different noise sources on the behavior of $R$ in view of the different formulations of the assembly activation event. The three characteristic setups we consider are as follows.
Figure \ref{Fig5}(a) refers to $R(D_1)$ dependence for the fixed very small $D_2$, whereas in Fig. \ref{Fig5}(b) is shown the $R(D_2)$ dependence for the fixed very small $D_1$. Finally, Fig. \ref{Fig5}(c) illustrates the behavior of $R(D_1)$ when $D_2$ is fixed at an intermediate value. Comparing Fig. \ref{Fig5}(a) and Fig. \ref{Fig5}(b), it stands out that the formulation $1$ on one hand, and the formulations $2$ and $3$ on the other hand, yield substantially different $R$ dependencies both below and above the stochastic bifurcation, cf. Fig. \ref{Fig3}(d). Below the bifurcation, the respectively smaller $R$ values obtained under formulation $1$ indicate that the corresponding activation process is more homogeneous compared to processes conforming to formulations $2$ and $3$. Furthermore, under formulations $2$ and $3$, the fluctuations around the mean $TFP$ expectedly increase in the noise domain that gives rise to stochastic bifurcation. The large $R$ values found there indicate strong irregularity of firing times over an ensemble of different stochastic realizations, cf. the peaks in Fig. \ref{Fig6}(a) and Fig. \ref{Fig6}(b). Sufficiently above the bifurcation, the $R$ values decrease for all three formulations. Nevertheless, in the stochastically supercritical state, the physical picture regarding the three formulations is in a sense reversed compared to the subcritical state. The deviations from the mean $TFP$ with $D_1$ are the largest for the activation process conforming to formulation $1$, whereas $R$ values for formulations $2$ and $3$ are smaller and fairly insensitive to noise increase, at least for sufficient $D_1$. An interesting remark is that by adhering to formulation $1$, as opposed to formulations $2$ and $3$, the coefficient of variation never crosses $1$, the value characteristic for the exponential distribution of events. It is well known that the latter distribution is consistent with the Poisson process \cite{HE05,DA01}.

Additional information on $R$ behavior can be gained by comparing the $TFP$ distributions of activation events over
an ensemble of different stochastic realizations for $(D_1,D_2)$ representative of the domains below or above the stochastic bifurcation (not shown). The main point concerns how likely are the deviations biased toward the shorter or longer activation times for an ensemble of realizations. Below the bifurcation, the ensemble of events under formulations $2$ and $3$ typically splits into fractions with short or very long $TFP$s. Though the former fraction is considerably larger, the long events still strongly influence the mean $TFP$. The partition into two fractions persists under formulation $1$, though the difference in duration between the two characteristic types of events is considerably reduced. The distribution of $TFP$s is then found to be bimodal, showing two relatively even peaks. These points are consistent with the differences displayed by the three $R(D_1,D_2)$ dependencies in Fig.\ref{Fig5} in the stochastically subcritical state. Nevertheless, above the bifurcation, qualitatively similar behavior for all three formulations of activation event is recovered. As expected, the $TFP$s follow a unimodal distribution centered around the short events, whereby a longer tail becomes visible for formulation $1$ if $D_1$ is increased. The latter accounts for the rise of $R$ with $D_1$ in Fig. \ref{Fig5}(c).

\subsection{Distribution of single units' $TFP$s}
\label{indtfps}

\begin{figure}
\centerline{\epsfig{file=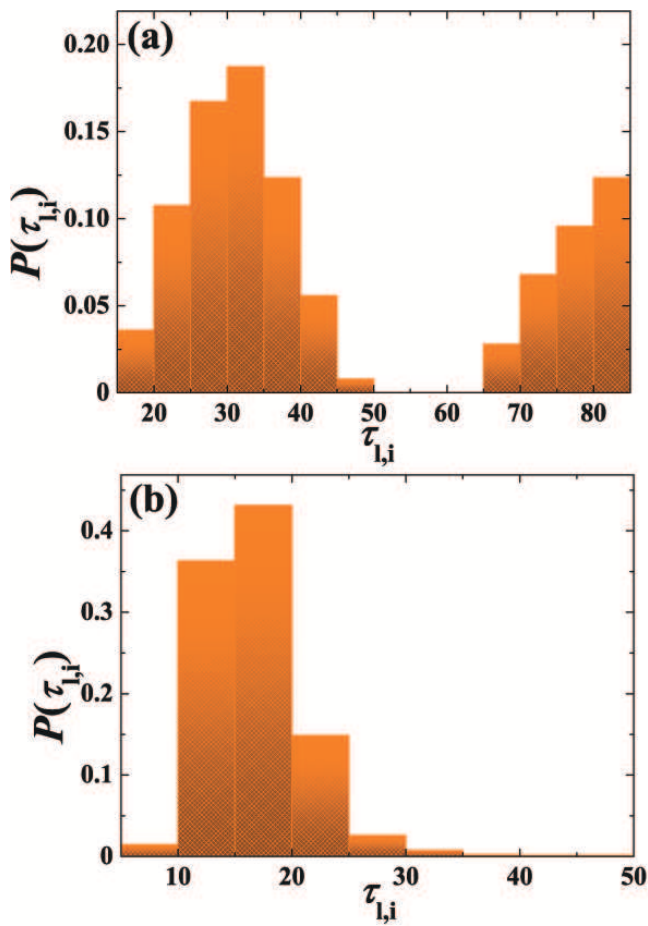,width=8.5cm}}
\caption{(Color online) Focus on local activation processes. (a) and (b) show typical distributions $P(\tau_{l,i})$ of individual unit $TFP$s $\tau_{l,i}$ for a single realization of the assembly activation process
at $(D_1,D_2)$ below and above the stochastic bifurcation, respectively. In the subcritical state, one finds two excitation waves, the first being induced by noise, whereas the other is due to relaxation of units. In the supercritical state, the units are primarily elicited by noise. The system parameters are
$D_1=0.,D_2=0.00009,c=0.1,N=300$ in (a) and $D_1=0.0002,D_2=0.00025,c=0.1,N=300$ in (b).} \label{Fig6}
\end{figure}

Having analyzed the properties of collective activation process for an ensemble of different stochastic realizations at fixed $(D_1,D_2)$, we now consider the dynamics of local activation processes for a \emph{single} realization of an \emph{assembly} event. In particular, we focus on how the $TFP$s of \emph{individual} units
are distributed for noise domains below or above the stochastic bifurcation. As for the underlying mechanism, an interesting point is to examine the contribution of noise-induced local activation events relative to the ones elicited by the relaxation processes of other units. We find that the typical profiles of the distribution  of individual $TFP$s below and above the stochastic bifurcation are quite different, as illustrated in Fig. \ref{Fig6}(a) and Fig. \ref{Fig6}(b), respectively. Note that the qualitatively analogous results are obtained if one compares the corresponding distributions of individual $TFP$s for all units over an ensemble of different stochastic realizations (not shown).

The bimodal distribution in Fig. \ref{Fig6}(a) suggests the existence of two "waves" of local excitation. While the first wave (short $TFP$s) is elicited by noise, the second wave is evoked by relaxation of the units that have already emitted the spike (such stimuli are conveyed via the interaction terms). This physical picture is reminiscent of the one obtained when determining the phase response curve of an assembly of oscillators subjected to random perturbations \cite{LP10}. Note that the smallest $TFP$s are around $20$ a. u., which is about the length of the spike excitation loop, i. e. the duration of the typical relaxation process. This indicates the existence of activation barrier associated to first pulse emission process of single units. Apart from the time scales involved, a remark should also be added regarding the sizes of waves, viz. the fractions of units participating the waves. The sizes are found to depend on $D_1,D_2$ and $c$, whereby the stronger $c$ expectedly amplifies the secondary wave. One would further expect the secondary wave to be substantially influenced by the topology of interactions within the network.

Above the stochastic bifurcation, the local activation events no longer maintain the two-wave-like form, see Fig. \ref{Fig6}(b). Instead, the typical distribution of single unit $TFP$s becomes unimodal, with some differences observed for the cases of intermediate vs large noise intensities. In particular, in the latter case the peak is more pronounced than in the former, whereby the distribution profile itself may be fitted to Lorentzian like-form. The very fact that the distribution of local $TFP$s is unimodal in the stochastically supercritical state implies that the activation of single units is almost exclusively driven by noise. Nonetheless, the point that the $TFP$s are typically of the order of the excitation loop suggests the lack of activation barrier in the first pulse emission processes of single units.

\subsection{Activation mechanisms below and above stochastic bifurcation}
\label{actmech}

While discussing $\tau(D_1,D_2)$ behavior in Fig. \ref{Fig3}, it has already been established that below the stochastic bifurcation, the assembly activation process under formulations $2$ and $3$ is not primarily affected by the first pulses of individual units, but is more likely contributed by the latter spikes. In these cases, we have identified gradual coherence build-up between individual spikes as the mechanism underlying assembly activation. Nevertheless, results in Fig. \ref{Fig6}(b) suggest that a potentially different scenario of assembly activation is typical above the stochastic bifurcation. Therefore, our next goal is to explicitly demonstrate the distinction between the mechanisms guiding the assembly activation below or above the stochastic bifurcation. To this end, we numerically determine the most probable trajectories for the first pulse emission process in the $(X,Y)$ configuration space and analyze the differences between the typical paths obtained for $(D_1,D_2)$ below or above the stochastic bifurcation, cf. Fig. \ref{Fig7}. The trajectories presented conform to formulation $2$ of the assembly activation event with the fixed threshold value $X_0=0.3$.

The details of the applied numerical method are as follows. For the given $(D_1,D_2)$, we consider the ensemble of fluctuation paths that start at moment $t_i$ from the deterministic fixed point $(X_{eq},Y_{eq})$ and reach the terminating boundary defined by the threshold $X_0$, cf. the solid line in Fig. \ref{Fig7}. At $t_i$, all the individual units are prepared to the same initial conditions, i. e. $(x(t_i),y(t_i))=(X_{eq},Y_{eq})$. The terminating time $t_f$, as well as $Y(t_f)$ are left unspecified. For the described ensemble, we consider statistics of the $(X(t),Y(t))$ position of trajectories as a function of time $t_i<t<t_f$ preceding the arrival to $X_0=0.3$. 
Naturally, the recorded trajectories may have quite distinct $t_f$ times. The proper statistical quantity to characterize such an assembly of paths in configuration space is the prehistory probability density, first introduced in \cite{DMSSS92} and successfully applied many times since \cite{NBK13}. The former can effectively be obtained if the time when each stochastic realization terminates is set to $t=0$, such that the behavior of the process during the initiation of the pulse is observed by looking backward in time. The appropriate prehistory probability distribution is then defined as $H(X,Y,t)dXdY=Pr[X(t)\in(X,X+dX),Y(t)\in(Y,Y+dY)|X(t_f)=X_0,X(t_i)=X_{eq},Y(t_i)=Y_{eq}], t_i<t<t_f, X<X_0$.
The most probable path for the first pulse emission process is determined by collecting the points $(X_m(t),Y_m(t))$ that correspond to the maximum of $H(X,Y,t)$ at the given $t$.

\begin{figure}
\centerline{\epsfig{file=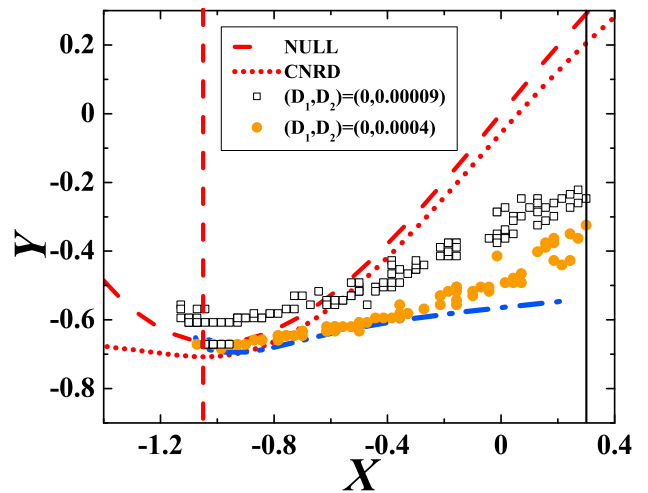,width=8.5cm}}
\caption{(Color online) Most probable activation paths for the global variables $(X,Y)$ under formulation $2$ of the activation event. The two illustrated profiles are typical for noise domains below (empty squares) and above (filled circles) the stochastic bifurcation. The shown paths refer to particular noise intensities $(D_1,D_2)=(0,0.00009)$ and $(D_1,D_2)=(0,0.0004)$, respectively. Note that the assembly's maximum likelihood trajectories correspond to the peak of the histogram obtained for an ensemble of different stochastic realizations as a function of time. The paths for the assembly are compared against the one for the approximate model \eqref{eq45} ("macroscopic $FHN$ unit"). The nullclines ($NULL$) of this model are presented by the dashed lines, whereas the canard-like trajectory ($CNRD$) is denoted by the dotted line. The trajectory generated by the effective Hamiltonian system associated with \eqref{eq45} is given by the dash-dotted line. The solid line indicates the threshold $X_0=0.3$. The results for the assembly refer to system size $N=100$ and are obtained by averaging over a $1000$ different stochastic realizations of the first pulse emission process.}
\label{Fig7}
\end{figure}

In Fig. \ref{Fig7}, the most probable activation paths typical for the noise domains below (above) the stochastic bifurcation are indicated by the empty squares (filled circles). In order to explain the differences between the presented paths, it is useful to invoke the approximate model \eqref{eq45}, which effectively assumes that all the units are strongly synchronized. In other words, that model refers to scenario where the assembly acts as a macroscopic $FHN$ unit. The dashed lines in Fig. \ref{Fig7} denote the nullclines of \eqref{eq45}. Further, since the latter has the form analogous to the dynamics of a single $FHN$ unit, we may borrow from our previous paper \cite{FTPVB14} the results obtained for such a setup. In particular, we plot the canard-like trajectory belonging to the "ghost-separatrix" \cite{KPLM13}, cf. the dotted line in Fig. \ref{Fig7}, as well as the trajectory generated by the system of effective Hamiltonian equations associated to \eqref{eq45} as discussed in our previous paper \cite{FTPVB14} (see the dash-dotted line in Fig. \ref{Fig7}). Note that the most probable activation paths for a single $FHN$ unit are only sensitive to the external/internal character of noise, but not to the particular noise intensities \cite{KPLM13,FTPVB14}.

For the trajectory corresponding to the noise domain below the stochastic bifurcation (empty squares) one finds that the approximate model \eqref{eq45} does not apply. From the latter point, one further infers that the units are not coherent during the initial part of the first pulse emission process, which is consistent with the already described scenario of gradual build-up of spike coherence. The mechanism of first pulse emission is quite different for the noise domain above the stochastic bifurcation. Recall that the stochastic bifurcation underlies transition to noise-induced oscillations, which are facilitated by the (stochastic) synchronization of individual units. With this in mind, it is expected to find that the corresponding most probable first pulse emission trajectory (filled circles) conforms much better to physical picture behind model \eqref{eq45}. In the initial part of the activation trajectory, there is (stochastic) synchronization between the units of the assembly, such that the assembly's activation path indeed fits well to the path given by the "macroscopic $FHN$ unit" model \eqref{eq45}. Note that the deviation from the path corresponding to \eqref{eq45} is observed in the latter part of the assembly activation trajectory. This is a natural consequence of the fact that the spike times of single units in the assembly are synchronized \emph{approximately}, but are not exactly matched. Thus, the effect of noise in the latter part of activation trajectory is felt more strongly for an assembly than in case of the effectively single $FHN$ unit model \eqref{eq45}. As expected, the described physical picture on the supercritical state breaks down if the noise intensity is too large.

\subsection{Mean $TFP$ anti-resonance as a system size effect}
\label{Size}

\begin{figure}
\centerline{\epsfig{file=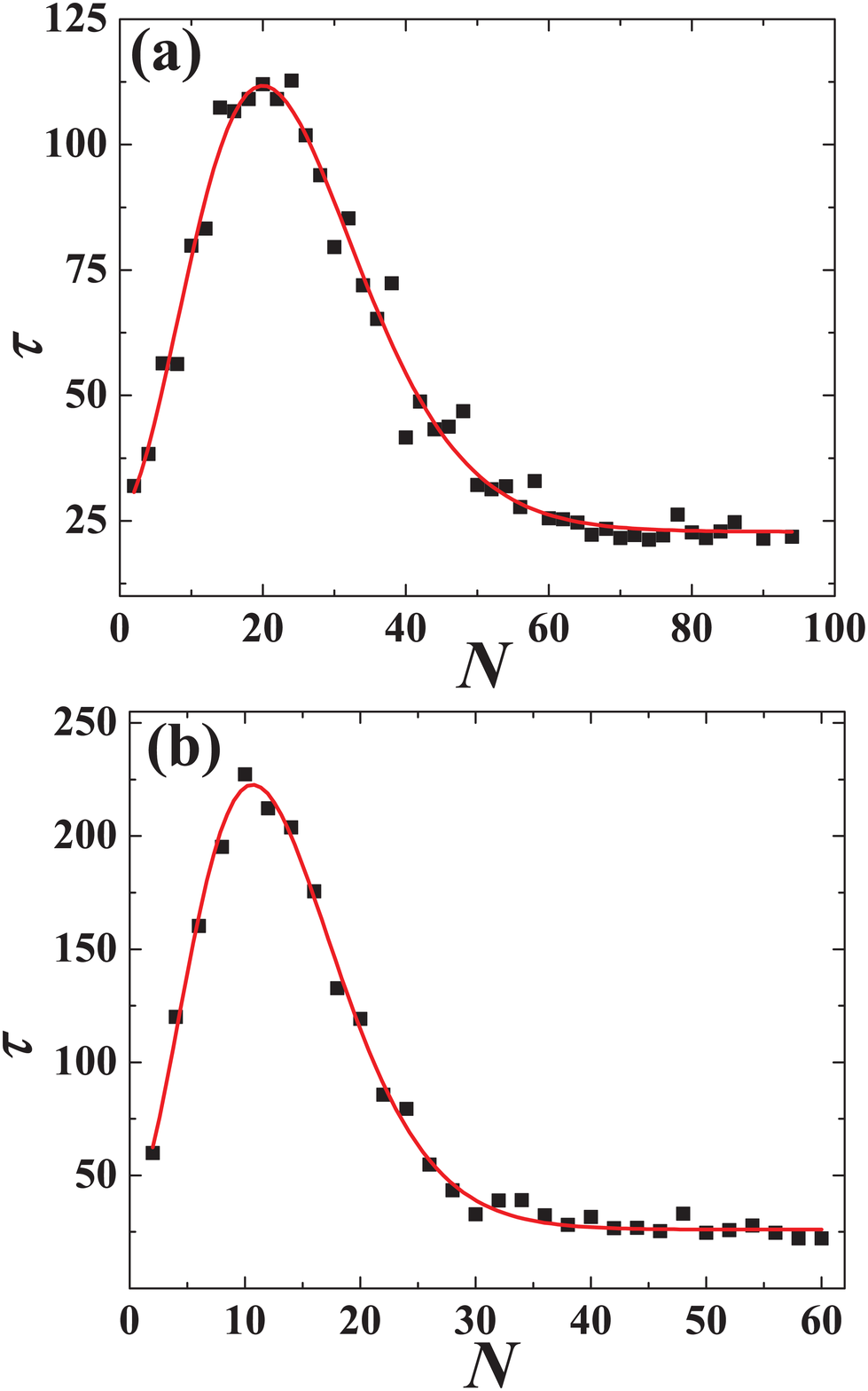,width=8.5cm}}
\caption{(Color online) Focus on the anti-resonant system-size effect for $\tau$. (a) and (b) illustrate the $\tau(N)$ dependencies typical for subcritical noise intensities, whereby (a) conforms to formulation $2$ and (b) to formulation $3$ of the assembly activation event. Both plots are obtained for the same set of parameter values $(D_1,D_2,c)=(0.0001365,0.0002255,0.1)$, with the threshold in (a) fixed to $X_0=0.4$. The stochastic averaging at each point has been carried out over an ensemble of $1000$ different activation paths.}
\label{Fig8}
\end{figure}

In this subsection, we briefly report on a system-size effect which discriminates between the different formulations of the assembly activation event. In particular, the effect is observed when applying formulations $2$ and $3$ which concern the dynamics of global variables, but is absent for formulation $1$ that refers to local activation events. The system size effect consists in the appearance of an anti-resonant peak in the dependence of mean $TFP$ with $N$ for the fixed parameter set $(D_1,D_2,c)$. The term anti-resonance is applied in a sense that under variation of $N$, the mean $TFP$s exhibit a maximum for the particular size of the assembly. Note that the effect is found for $(D_1,D_2)$ values both below and above the stochastic bifurcation, but is substantially more pronounced in the stochastically subcritical state. As an illustration, in Figs. \ref{Fig8}(a) and \ref{Fig8}(b) the $\tau(N)$ dependencies are shown corresponding to formulations $2$ and $3$, respectively, whereby the noise intensities are typical for the domain below the stochastic bifurcation. Apart from the fact that the formulation $1$ does not admit similar behavior, it also yields $\tau(N)$ dependence that exhibits only a weak decline with $N$. As already indicated, under formulations $2$ and $3$ the anti-resonant effect persists above the stochastic bifurcation. Nevertheless, the $\tau(N)$ maximum in this domain reduces with noise and further shifts to smaller $N$. For a comprehensive understanding of the described anti-resonant system size effect, one would have to carry out an elaborate analysis on a number of issues independent on the present study. The focus should primarily lie with the mechanisms contributing the long activation events, including the details on how the competition between the noise- and relaxation-induced local activation processes is affected by the system size.

\section{Summary}
\label{Conc}

The present study has been aimed at characterizing the excitable behavior and the noise influenced activation process of an assembly of class II excitable units whereby each unit is subjected to external and internal noise. The analysis is an extension of our previous work \cite{FTPVB14} concerning the activation processes for a single and two interacting $FHN$ elements. In conceptual terms, three points can be regarded as most relevant ones. First, we have explicitly demonstrated that an assembly of excitable $FHN$ units may itself exhibit macroscopic excitable behavior. As a second point, we have established the qualitative analogy between the statistical properties of the noise-driven activation processes for a single $FHN$ unit and an assembly. Finally, it has been shown that depending on the noise intensities, two qualitatively distinct local mechanisms may have prevailing effect on the assembly activation process.

The first point has been achieved by implementing the appropriate $MF$ approach, which ultimately yields the approximate model \eqref{eq4}. As in case of a single $FHN$ unit, the threshold-like dynamics separating the assembly's spiking responses from the small-amplitude excitations has been linked to the ghost-separatrix. An additional result is that compared to excitable dynamics of a single $FHN$ unit for the analogous parameter set,
the dynamics of the assembly exhibits larger sensitivity to initial conditions in vicinity of the ghost-separatrix.

As for the second point, establishing the qualitative analogy in statistical properties of the activation processes of a single unit and an assembly is a non-trivial result that could not be expected \emph{a priori}. To facilitate the comparison between the respective activation processes, we have introduced three alternative formulations of the assembly activation event, each emphasizing different aspects of collective dynamics. Formulation $1$ deliberately makes no mention of global variables, such that the effects of coherence of individual units' spikes are left aside. At variance with this, formulations $2$ and $3$ are stated in terms of global variables, such that the former involves an explicit scalar threshold for the spiking response, whereas the latter incorporates an appropriate terminating boundary set analogous to the one we adopted for a single and two interacting $FHN$ units \cite{FTPVB14}.

The analysis on statistical properties of assembly activation process has been carried out by determining the $\tau(D_1,D_2)$ and $R(D_1,D_2)$ dependencies for the three alternative formulations of the activation event.
All three formulations yield qualitatively analogous results for the mean $TFP$s, but certain quantitative differences are manifested in the subcritical regime. These differences are due to microscopic mechanisms behind the assembly activation, and concern the role of spike coherence in the onset of activation event. In particular, the activation processes under formulations $2$ and $3$ are mediated by the gradual build-up of spike coherence, whereas in case of formulation $1$ spike coherence is not relevant.

We have further found that the statistical properties associated to macroscopic excitable behavior qualitatively resemble those for a single unit \cite{FTPVB14}. The observed qualitative similarity, both with respect to three formulations of the assembly activation event and compared to the case of a single $FHN$ unit, has been tied to stochastic bifurcation that underlies transition from the stochastically stable fixed point to continuous oscillations. The stochastic bifurcation has been analyzed within the framework of the $MF$ model \eqref{eq4}, demonstrating that the latter undergoes Hopf bifurcation for the noise intensities that qualitatively coincide with those that give rise to stochastic bifurcation of the exact system. The stochastic bifurcation has been shown to account for the transition between the $(D_1,D_2)$ domains admitting large mean $TFP$s and the plateau region.

As for our third main point, the analysis of the mechanisms behind the local activation processes has revealed substantial differences between the stochastically subcritical and the supercritical state. In the former case, two excitation waves may be discerned: the first is elicited by noise, whereas the second wave is due to relaxation of units, viz. is evoked by the interaction terms. For this scenario, one is able to confirm the existence of a potential barrier associated to single unit activation. At variance with this, above the stochastic bifurcation, the local activation processes are strongly influenced by noise.

The respective roles of noise and interaction terms in the mechanisms leading to assembly activation have further been examined by considering the most probable activation paths characteristic for the noise domains below and above the stochastic bifurcation. The analysis is based on comparing the results for the assembly to the appropriate most probable activation path of the approximate model \eqref{eq45}. The latter model assumes strong synchronization between the units, effectively describing the assembly as a macroscopic $FHN$ unit. For the subcritical state, the analysis corroborates the gradual build-up of spike coherence as the leading mechanism behind the assembly activation. This mechanism is naturally influenced by the interaction terms. Nevertheless, above the stochastic bifurcation, noise facilitates stochastic synchronization between the units, rendering the interaction terms negligible. In this scenario, the assembly activation process may indeed be compared to that of the approximate model \eqref{eq45}.

In terms of details specific for the assembly excitable behavior, an interesting finding concerns a system-size effect, which consists in the appearance of an anti-resonant peak in the $\tau(N)$ dependence under fixed $(D_1,D_2)$. While $\tau(N)$ displays a maximum for arbitrary noise intensities, the maximum is still substantially more pronounced below than above the stochastic bifurcation.

Apart from providing insights into the assembly activation process, the present study has raised a number of novel issues. For example, the future research may include a systematic study on the mechanisms behind the reported system-size effect, or could focus on the impact of connection topology on the activation process. Another relevant point would be to consider whether the qualitative aspects of behavior found here are paradigmatic, i. e. whether  they persist if the assembly is built of type I instead of type II excitable units.

\begin{acknowledgments}
This work has been supported by the Ministry of Education, Science and Technological Development of the Republic of Serbia, under project Nos. $171017$ and $176016$. M.P. acknowledges support from the Slovenian Research Agency (Grant P5-0027), and from the Deanship of Scientific Research, King Abdulaziz University (Grant 76-130-35-HiCi).
\end{acknowledgments}

\end{document}